\documentclass[12pt,preprint]{aastex}
\begin{document}
\title{A Comprehensive Study of Gamma-Ray Burst Optical Emission: III. Brightness Distributions and Luminosity Functions of Optical Afterglows}
\author{Xiang-Gao Wang\altaffilmark{1}, En-Wei Liang\altaffilmark{1,2}, Liang Li\altaffilmark{1}, Rui-Jing Lu\altaffilmark{1}, Jian-Yan Wei\altaffilmark{2}, Bing Zhang\altaffilmark{3,1}}

\altaffiltext{1}{Department of Physics and GXU-NAOC Center for Astrophysics and Space Sciences, Guangxi University, Nanning 530004,
China;lew@gxu.edu.cn}
\altaffiltext{2}{National Astronomical Observatories, Chinese Academy of Sciences, Beijing, 100012, China}
\altaffiltext{3}{Department of Physics and Astronomy, University of Nevada, Las Vegas, NV 89154,USA}

\begin{abstract}
We continue our systematic statistical study on optical afterglow
data of gamma-ray bursts (GRBs). We present the apparent magnitude
distributions of early optical afterglows at different epochs ($t =
10^2$ s, $t = 10^3$ s, and 1 hour) for the optical lightcurves of a
sample of 93 GRBs (the global sample), and for sub-samples with an
afterglow onset bump or a shallow decay segment. For the onset
sample and shallow decay sample we also present the brightness
distribution at the peak time $t_{\rm p}$ and break time $t_{\rm
b}$, respectively. All the distributions can be fit with Gaussian
functions. We further perform Monte Carlo simulations to infer the
luminosity function of GRB optical emission at the rest-frame time
$10^3$ seconds, $t_{\rm p}$, and $t_{\rm b}$, respectively. Our
results show that a single power-law luminosity function is adequate
to model the data, with indices $-1.40\pm 0.10$, $-1.06\pm 0.16$,
and $-1.54\pm 0.22$, respectively. Based on the derived rest-frame
$10^3$ s luminosity function, we generate the intrinsic distribution
of the R-band apparent magnitude $M_{\rm R}$ at the observed time
$10^{3}$ seconds post trigger, which peaks at $M_{\rm R}=22.5$ mag.
The fraction of GRBs whose R-band magnitude is fainter than 22 mag,
and 25 mag and at the observer time $10^3$ seconds are $\sim 63\%$
and $\sim 25\%$, respectively. The detection probabilities of the
optical afterglows with ground-based robotic telescopes and UVOT
onboard {\em Swift} are roughly consistent with that inferred from
this intrinsic $M_{\rm R}$ distribution, indicating that the
variations of the dark GRB fraction among the samples with different
telescopes may be due to the observational selection effect,
although the existence of an intrinsically dark GRB population
cannot be ruled out.
\end{abstract}

\section{Introduction\label{sec:intro}}
Optical afterglows have been detected from $\sim 250$ gamma-ray
bursts (GRBs) since the first optical counterpart was identified in
February 28, 1997 (van Paradijs et al. 1997). The observed magnitudes
span from 5.5 to 28 mag, in a time-window from tens to $10^7$
seconds past the GRB trigger, with telescopes whose apertures range
from tens of centermeters to 10 meters. Statistical studies of
optical afterglow lightcurves have been carried out by some authors
and some general features of the lightcurves have been reported. For
example, optical lightcurves have been compiled in the rest frame of
GRBs, and two universal tracks have been claimed (Liang \& Zhang
2006; Nardini et al. 2006; Kann et al. 2006. c.f., Melandri et al.
2008, Oates et al., 2009, Zaninoni et al 2013). Panaitescu \&
Vestrand (2008, 2011) showed some general features of the early
bumps and plateaus in the optical lightcurves. Akelof \& Swan (2007)
estimated the apparent optical brightness distribution function and
suggested that the apparent optical magnitude distribution peaks at
$R\sim 19.5$. Kann et al. (2010, 2011) compared the optical
lightcurves of two types of GRBs. We carry out a systematical
analysis of the optical data of GRBs and present our results in a
series of papers. In the first two papers of this series (Li et al.
2012, paper I; Liang et al. 2012, paper II), we presented general
features of a ``synthetic'' optical lightcurve based on our
decomposition analysis of lightcurves, and focused on the
statistical properties and the physical implications of the optical
flares, the shallow decay segment, the afterglow onset bump, and the
late re-brightening component. We showed that the optical flares and
the shallow decay segment may signal late central engine activity
(Paper I), and that the onset bumps and late re-brightening bumps
may probe the properties of the fireball and its surrounding medium
density (Paper II; see also Rykoff et al.2009; Oates et al. 2009;
Liang et al. 2010; L\"{u} et al. 2012; Ghirlanda et al. 2012; Yi et
al. 2012).

The luminosity function of GRB afterglows is poorly known since no
complete sample in a given threshold is available. Observationally,
the detection of an optical counterpart of a GRB depends on the
instrument, exposure time, observation epoch, etc. For example, the
UV-optical telescope (UVOT) on board {\em Swift} promptly slews to
the GRB trigger positions, and $\sim 27\%$ of GRBs were detected  at
the $3\sigma$ level (in an individual exposure; Roming et al. 2009),
but $\sim 60\%$ of all GRBs have been detected by ground-based
telescopes in the spectral bands redder than UVOT (e.g., Akelof \&
Swan 2007 using GCN data). The optical afterglows may also suffer
significant dust extinction by the host galaxies of GRBs, since long
GRBs are believed to be born in dusty, opaque star-forming regions
(Reichart \& Price 2002; Klose et al. 2003; Vergani et al. 2004;
Levan et al. 2006). They are not visible at all in the optical band
due to the Lyman-$\alpha$ absorption of neutral hydrogen, if a GRB
originates at a high redshift (Jakobsson et al. 2004; 2005). In
addition, the observed optical emission may be the superposition of
multiple components with distinct physical origins, whose strengths
and decay slopes may vary from burst to burst (e.g., Li et al.
2012). These effects cause complications in revealing the luminosity
function of optical afterglows.

This paper continues our systematic statistical study on optical
afterglow data. We study a sample of 93 GRBs with optical detections
before $t<1$ hour (the global sample), and two sub-samples whose
lightcurves show an onset bump or a shallow decay segment,
respectively (see e.g. Paper I and II). For these three samples, we
present the apparent R-band magnitude distributions of early optical
afterglows at several different epochs ($t = 10^2$ s, $t = 10^3$ s
and 1 h) after the triggers. We then use Monte Carlo simulations to
investigate the intrinsic luminosity function of the GRB optical
afterglows at the rest-frame $t = 10^3$ s, at the peak time $t_p$
for the sample that show an onset bump in the lightcurves, and at
the break time $t_p$ for the sample that show a shallow decay
segment. The early optical lightcurves are rich in features that may
be attributed to different physical origins. A significant fraction
of the lightcurves are dominated by a clear onset bump that signals
the deceleration of the fireball, which is most sensitively defined
by the fireball initial Lorentz factor (Paper II). Some others show
a shallow decay segment, which is related continuous energy
injection into the blastwave (Paper I). The end time of this phase
$t_b$ is related to a critical time of late energy injection from
the GRB central engine or related to a critical time when pile-up of
flare materials onto the blastwave is over. We therefore also study
the luminosity functions at $t_p$ and $t_b$, which may hold the key
in studying the fireball parameters and the late central engine
activities. We describe our samples in \S 2. The observed R-band
magnitude distributions at various epochs are presented in \S 3. A
Monte Carlo simulation analysis is presented in \S 4 to infer the
optical luminosity function at rest-frame $t = 10^3$ s, $t_p$ and
$t_b$. Conclusions and discussion are presented in \S 5.

\section{Sample and Data\label{sec:data}}
We have complied a large sample of GRB optical data from published papers
or from GCN Circulars when no published paper is available. Well-sampled lightcurves are available for 146 GRBs, as shown in Figure \ref{total}. Galactic extinction correction is made to the data by using a reddening map presented by Schlegel et al. (1998). Most data are in the R-band. Those data that were obtained in other wavelengths are corrected to the R band with the optical spectral indices ($\beta_{\rm O}$) collected from the literature \footnote{The convention $F_\nu \propto \nu^{-\beta_{\rm O}}$ is
used. An optical spectral index $\beta_O=0.75$ is adopted for those
GRBs whose $\beta_O$ is not available. The extinction $A_{\rm V}$ of
the host galaxy and the spectral index of the optical afterglow for
each burst from the same literature is also used to reduce the
uncertainties introduced by different authors.}.  These lightcurves
were decomposed into multiple empirical components with smooth
broken power-law functions. The details of the fitting results are
reported in Papers I and II. From 146 GRBs, we focus on early
optical afterglows and select 93 GRBs that have optical detections
at $t<1$ hour. This is our {\em global} sample. We calculate the
brightness at $t=10^2, 10^3$ seconds and 1 hour for this sample
using empirical fits to the lightcurves. The redshift, temporal
coverage of the lightcurve data, and the derived brightness in these
epochs are reported in Table 1. A sub-sample of 39 GRBs have a
smooth afterglow onset bump in the lightcurve, and a sub-sample of 26
GRBs have a shallow decay segment in the lightcurves (Liang et al.
2012). We also derive the brightness at $t=10^2, 10^3$ seconds and 1
hour for these two sub-samples based on our empirical fits available
in Papers 1 and II. The brightness at the peak time ($t_{\rm p}$) of
the afterglow onset bumps and the end time ($t_{\rm b}$) of the
shallow decay segments are also calculated with the results of the
empirical fits. They are reported in Tables 2 and 3.

Note that the current deepest survey for early optical
afterglows is made with the GROND telescope (Greiner et al. 2011). It
has a limit of $M_{\rm R} \sim 25$ mag. As shown in  Figure \ref{total},
we do not find GRBs that have detections dimmer than 22 mag at
$t<10^2$ seconds in our sample. We check the observations with the
GROND telescope for $t<10^3$ seconds and found that some GRBs have
been detected with $R>22$ mag for a few cases (such as GRB 081029,
R=22.8 at 240 s, GCN 7231; GRB 081029, R $\sim$ 24.3 at 660s, GCN
8731). Our sample includes only the GRBs that have well-sampled
lightcurves. Therefore, the non-detection with $M_{\rm R}>22$ mag at
$10^3$ seconds in our sample is due to the sample selection effect.
This selection effect, however, does not impact on our analysis
results since we perform simulations with a limit of $M_{\rm R}=19$ mag
(see \S 4.2).

\section{Apparent Magnitude Distributions}
Among the 93 GRBs in our sample, 89 GRBs have optical detections
earlier than $10^3$ s, and 36 GRBs have optical detections earlier
than $t = 10^2$ seconds. We obtain the luminosity at t=$10^2$
seconds for the 36 GRBs from our empirical fits. We also derive the
luminosity at t=$10^2$ for the other GRBs that have optical
detection earlier than $10^3$ seconds (most of them have detection
very closed to $10^2$ seconds, see Figure 1) by extrapolating our
empirical fits to t=$10^2$ seconds. The lightcurves of 7 GRBs are
poorly sampled or rapidly increase between $10^2$ and $10^3$
seconds. We do not derive the luminosity at t=$10^2$ from these
lightcurves. We therefore finally obtain a sample of 82 GRBs that
have optical luminosity at t=$10^2$ seconds. Figure \ref{Dis_obs}
shows the distributions of the apparent magnitudes at $t=10^2, 10^3$
seconds and 1 hour post the GRB triggers for the global lightcurves.
The distributions at $t=10^2$ and $10^3$ seconds are well fit with a
Gaussian function, and the derived central values with 1 $\sigma$
deviations are $M_{\rm R,c}=16.1\pm 1.8$ mag and $17.3\pm 1.8$ mag
(1$\sigma$), respectively. The magnitudes at $t=1$ hour of the 93 GRBs are narrowly distributed in the range $M_{\rm R}=14-22$
mag, with a Gaussian fit of $M_{\rm R,
c}=18.4\pm 1.6$ mag. The fact that these distributions can be fitted
with Gaussian without a sharp cut-off at the low flux end may be due
to the fact that the current sample was obtained from observations
of telescopes with different flux limits.

A smooth onset hump is observed in 39 GRBs in our sample. We show
the brightness distributions for this component at $t=10^2$, $10^3 $
seconds, 1 hour and at $t=t_{\rm p}$. In general, they are roughly
consistent with that for the global lightcurves as shown in Figure
2. The brightness distribution at $t=t_{\rm p}$ is well fit with a
broad Gaussian function, i.e., $M_{\rm R, c}=15.6\pm 2.2$ mag, with
95\% having $M_{\rm R}<19$. This may
be due to the observational selection effect since the onset bump is
usually detected using small aperture telescopes in a moderate limit of
$M_{\rm R}\sim 19$ mag.

The standard fireball model suggests that the decay slope of the
afterglows should be steeper than 0.75 if no additional energy is
continuously injected into the blastwave. Liang et al. (2012) defined
a shallow decay segment with the criterion that the decay
slope is initially shallower than 0.75, which transits to a
steeper decay after a break time $t_b$. Such a segment is observed in 26
GRBs in our GRB sample.  We also show the brightness distributions
for this component at $t=10^2$, $10^3$ seconds, 1 hour and at $t=t_{\rm
b}$ in Figure \ref{Dis_obs}. They are also roughly consistent
with those for the global sample. This is reasonable since the
early optical lightcurves are usually dominated by the onset bump or
the shallow decay segment. The brightness at $t_{\rm b}$ is $M_{\rm
R, c}=18.3$ mag.

\section{CONSTRAINING THE INTRINSIC LUMINOSITY FUNCTIONS WITH
MC SIMULATIONS\label{sec:normal}}
The luminosity of a GRB afterglow at a given time depends on the kinetic energy, micro-physical parameters and radiation efficiency of the fireball (e.g., Sari et al. 1998). No universal values for these quantities are found among bursts, and the correlation between the luminonsites of the prompt gamma-rays and the optical afterglow is loose. One cannot simply infer the optical luminosity function from that of the prompt gamma-ray emission. Here we take optical afterglow as independent of their $\gamma$-ray luminosity function, and perform Monte Carlo simulations.
Since the optical luminosity depends on the epoch, in the following we constrain the luminosity function at
$10^3$ s after the GRB trigger in the rest frame of GRBs. In order to study the deceleration physics and energy injection physics, we also constrain the luminosity functions at $t_p$ and $t_b$, respectively. Since these times differ from burst to burst, there is no need to differentiate the rest-frame and the observer-frame.

\subsection{Model}
The number density of GRBs at redshift $z\sim z+dz$ is given by
\begin{equation}\label{density}
n(z)=\frac{dN}{dz}=\frac{R_{\rm
GRB}(z)}{1+z}\frac{dV(z)}{dz},
\end{equation}
where $R_{\rm GRB}(z)$ is the co-moving GRB rate as a function of
$z$, the factor $(1+z)^{-1}$ accounts for the cosmological time
dilation of the observed rate, and $dV(z)/dz$ is the co-moving
volume element. We assume $R_{\rm GRB}$ traces the star formation rate
and metallicity history, e.g. (Kistler et al. 2008; Li 2008;
Qin et al. 2010; Virgili et al. 2011; Lu et al. 2012)
\begin{equation}\label{rate}
R_{\rm GRB}(z)=\rho_{\rm0} R_{\rm SFR}(z)(1+z)^{\delta}\Theta(\epsilon,z),
\end{equation}
where $\rho_{\rm0}$ is the local GRB rate in units of Gpc$^{-3}$ yr$^{-1}$,
$(1+z)^{\delta}$ accounts for the possible GRB rate evolution effect
in excess of the SFR rate, $\Theta(\epsilon,z)$ is the fractional
mass density belonging to the metallicity below $\epsilon Z_{\odot}$
at a given $z$ ($Z_{\odot}$ is the solar metal abundance), and
$\epsilon$ is determined by the metallicity threshold for the
production of GRBs. The star-forming rate $R_{\rm SFR}(z)$ is taken
as (Langer \& Norman 2006; Kistler et al. 2008),
\begin{equation}
  R_{\rm SFR}(z) \propto \left\lbrace \begin{array}{ll}~(1+z)^{3.44}, ~~~~~~z \leq 1\\
                                              (1+z_{\rm peak})^{3.44}, ~~~~~~z > 1 \\
\end{array}. \right.
\label{SFR}
\end{equation}
The $\Theta(\epsilon,z)$ is parameterized as
(Hopkins \& Beacom 2006; Langer \& Norman 2006)
\begin{equation} \label{meta}
\Theta (\epsilon, z)= \frac{\hat{\Gamma}(\hat\alpha+2,\epsilon^{\hat\beta}
10^{0.15 \hat\beta z} )}{\Gamma(\hat\alpha+2)},
\end{equation}
where $\hat\alpha$ and $\hat\beta$ are the lower and higher end indices of the
galaxy mass - metallicity relation, and $\hat{\Gamma}(a,x)$ and $\Gamma(x)$ are
the incomplete and complete gamma function (Langer \& Norman 2006;
Kistler et al. 2008), respectively. We take $\alpha_\gamma=-1.16$, $\beta_\gamma=-2$,
$\epsilon=0.4$, and $\delta=0.4$
(Qin et al. 2010)\footnote{As discussed in Qin et al. (2010), the metallicity factor may vary from 0.2 to 0.6 derived from Monte Carlo simulations for the GRB rate. Being due to small samples in our analysis, we find that the variation of the fraction from 0.2 to 0.6 does not significantly change our results.}.

Assuming the optical spectrum as $F_\nu\propto \nu^{-\beta_{\rm O}}$, the observed flux by considering the k-correction effect would be
\begin{equation}\label{fobs}
\ F_\nu=\frac{L_\nu(\nu_0)(1+z)^{1-\beta_{\rm O}}}{4\pi D_{\rm L}^{2}(z)},
\end{equation}
where $D_{\rm L}(z)$ is the luminosity distance at {\em z}, $\nu_0$ is the frequency in the burst frame. The observed R-band magnitude by considering dust extinction in the GRB host galaxy ($A^{\rm host}_{\rm V}$) is
\begin{equation}\label{mag}
M_{\rm R} =\frac{\log(F_{\rm R}/F_{\rm R, 0})}{-0.4}+A^{\rm host}_{\rm V},
\end{equation}
where $F_{\rm R, 0}=1.218\times 10^{-5}$ erg cm$^{-2}$ s$^{-1}$. Here we take the central frequency and the full-width-half-maximum of the R band as $7000{\AA}$ and $2200{\AA}$, respectively. We assume that the optical afterglows are the synchrotron radiation in the spectral regime $\nu_m<\nu<\nu_c$, and take $\beta_{O}=0.75$ for an electron power-law index $p=2.3$.

\subsection{Simulation procedure and results}
We assume that the luminosity functions of GRB optical afterglow at rest-frame $t = 10^3$ s post trigger, at $t_p$ and $t_b$ can be all characterized as a single power-law or a smooth broken power-law\footnote{It is known that the GRB luminosity function can be characterized with a broken power-law or a cut-off power-law (e.g.,Schmidt 2001; Zhang et al. 2004; Guetta et al. 2005; Liang et al. 2007). Therefore, we try to use the same function form to describe the optical afterglow luminosity function.}
\begin{eqnarray}\label{LF}
\Phi(L) & = & \Phi_{0}(L/L_{0})^{\alpha}\\
\Phi(L) & = & \Phi_{0}[(\frac{L}{L_{\rm
b}})^{\alpha_{1}}+(\frac{L}{L_{\rm b}})^{\alpha_{2}}]^{-1},
\end{eqnarray}
where $\Phi_{0}$ is a normalization constant and $L_{0}$ is taken as $10^{46}$ erg/s. We use the optical afterglow data to constrain the luminosity function parameters by Monte Carlo simulations in the luminosity range of $[10^{43},10^{50}]$ erg s$^{-1}$. The details of our simulation procedure are as follows.

(1) Generate a redshift and a luminosity for the optical
afterglow from the probability distributions of the
redshift and luminosity based on Eqs. \ref{density} and \ref{LF} (or
8) with a set of free parameters, respectively. We take a redshift
range of [0, 6.3]\footnote{The observed R-band emission would
be significantly affected by the Lyman-$\alpha$ absorbtion for a GRB at
$z>6$. We therefore take an upper limit of the redshift as that of
GRB 050904, i.e., $z=6.3$.} and a luminosity range of
$[10^{43},10^{50}]$ erg s$^{-1}$.

(2) Calculate the observed apparent magnitude of the simulated GRB with
Eqs. (\ref{fobs}) and (\ref{mag}). The $A_{\rm V}$ value is
generated from the $A_{\rm V}$ distribution of current sample with
$A_{\rm V}$ available, which is shown in Table 4 and Figure
\ref{Av}. The distribution of $A_{\rm V}$ can be fitted with a
log-normal function, i.e., $\log(\emph{A}_{V})=-0.63\pm0.42$(1$\sigma$).

(3) Compare the mock sample with the observations in the $\log
M_{\rm R}-\log (1+z)$ plane. To constrain the $10^3$ s luminosity
function, for the observational data we identify the relevant epoch
in the observer frame to make sure that it corresponds to the
rest-frame $10^3$ s post trigger. The observed sample suffers
observational biases. The optical data in the current sample are
collected from observations with different telescopes. The optical
telescopes do not have a clean flux limit as high energy
instruments. The early optical data are usually observed with
robotic telescopes with small apertures. The flux limit of these
telescope is usually $19$ mag. We therefore compare the mock sample
with the observed sample by screening the samples with a threshold
of $M_{\rm R}=19$ mag. We evaluate the consistency of the $z$ and
$M_{\rm R}$ distributions between the mock and observed samples with
a Kolmogorov-Smirnov test (K-S test). Note that the magnitude of a simulated GRB is derived from the luminosity and redshift. To compare a simulated GRB sample with observations, the GRB sample is screened by the instrumental threshold. The magnitude of a GRB with larger luminosity at higher redshift may be the same as a GRB with smaller luminosity at lower redhisft. Therefore, the consistency of the magnitude distributions between the simulated and observed samples alone cannot ensure the consistency between the redshift distributions. Therefore, we define a global probability of the K-S test ($P_{\rm K-S}^{\rm G}$) to measure the consistency of both the magnitude and redshift distributions between the simulated and observed sample as $P_{\rm K-S}^{\rm
G}=P_{\rm K-S}^{\rm M}\times P_{\rm K-S}^{z}$, where $P_{\rm
K-S}^{\rm M}$ and $P_{\rm K-S}^{\rm z}$ are the probabilities of the
K-S tests for the $M_{\rm R}$ and $z$ distributions, respectively. A
larger value of $P_{\rm K-S}$ indicates a better consistency.
Generally, a value of $P_{\rm K-S}$ $>$0.1 is acceptable to claim
statistical consistency, whereas a value of $P_{\rm K-S}<10^{-4}$
convincingly rejects the consistency hypothesis.

We repeat the above steps to search for the best consistency in a
broad parameter space. In our simulations, a mock GRB is
characterized with a set of $\{z, L, A_{\rm V}\}$. We generate a
sample of $10^5$ mock GRBs based on a luminosity function with a
given power-law index $\alpha$ (or a set of $\{\alpha_1, \alpha_2\}$
for a broken luminosity function). We then measure the consistency
between the mock GRB sample and the observed sample with the K-S
test. We make $10^5$ trials for $\alpha$ (or $\{\alpha_1, \alpha_2
\}$ for a broken luminosity function) randomly in the range of
[0,-2.5] and show $P_{\rm K-S}^{\rm G}$ as a function of $\alpha$ to
find the $\alpha$ value that gives the best consistency between the
simulations and observations. We evaluate the luminosity function of
the global lightcurves at $t=10^3$ seconds. The selection of this
specified epoch is due to that both ground-based and spaced-based
telescopes with different apertures may make observations at this
epoch. We also estimate the luminosity functions at the afterglow
onset peak and at the break time of the shallow decay segments. We
find that the constraints on the parameters of a broken power-law
luminosity function with the current sample is loose, and a single
power-law luminosity function is adequate to model the current
sample. The $P^{\rm G}_{\rm K-S}$ distributions for different
$\alpha$ values is shown in Figure \ref{Pks_alpha_dis}. The Gaussian
fits yield $\alpha=-1.40\pm 0.10$, $-1.06\pm 0.16$, and $-1.54\pm
0.22$ for the luminosity functions at rest-frame $t=10^3$ seconds of
the global-sample lightcurves, at the peak time $t_{\rm p}$ of the
onset bumps, and at the break time $t_{\rm b}$ of the shallow decay
segments, respectively. The errors are in 1 $\sigma$ confidence
level. We also randomly generate a sub-sample that has the same size
as the observed sample based on a luminosity function with
$\alpha=-1.40\pm 0.10$. Figures \ref{MCtotal1000}-\ref{MCshallow}
demonstrates the consistency between the simulated and observational
samples in the 2-dimensional $M_{\rm R}-\log (1+z)$ plane and in the
1 dimensional $M_{\rm R}$ and $\log (1+z)$ distributions. One can
observe that the luminosity functions can produce the observational
data. The slope of the luminosity functions for the global sample
lightcurves and that of the shallow decay segment break time $t_{\rm
b}$ are consistent. This would be due to the fact that $t_{\rm b}$ is usually
at several hundreds to several thousands of seconds with a typical
value of $\sim 10^{3}$ seconds (Li et al. 2012). The slope of the
luminosity function at the onset bump $t_{\rm p}$ is shallower than
that derived at $t_{\rm b}$.

Based on the derived luminosity function, we generate the intrinsic
distribution of $M_{\rm R}$ at the observed $10^{3}$ seconds for the
global sample. We generate a sample of $10^{5}$ GRBs in the
luminosity range [$10^{43}\sim 10^{50}$] erg/s based on the
rest-frame luminosity function with $\alpha=-1.40\pm 0.10$. Fixing
the model parameters of the GRB rate (Eq. \ref{rate}), we calculate
the $M_{\rm R}$ values of the mock GRBs at the observed $10^3$ s
post-trigger. To do so, we extrapolated lightcurves from the
rest-frame $10^3$ s to the observer-frame $10^3$ s by introducing a
decay slope for each lightcurve, the distribution of which is
derived from the observed decay-slope distribution. The resulting
observer-frame $10^3$ s $M_{\rm R}$ distribution is shown in Figure
\ref{MCbrightness}. The accumulated probability distribution of
$M_{\rm R}$ is also shown. The peak of the distribution is $M_{\rm
R}=22.5$ mag, which is dimmer by $3$ mag than the peak of the
apparent optical magnitude distribution ($M_{\rm R}\sim 19.5$)
reported by Akelof \& Swan (2007). This would be due to
observational selection effect in the sample used by Akelof \& Swan
(2007).

We examine the detection probability for GRB optical afterglow
surveys using telescopes with different flux limits. Small aperture
robotic telescopes for early optical afterglow observations, such as
ROTSE (Akerlof et al. 2003) and KAIT (Filippenko et al. 2001),
usually have a limit of $M_{\rm R}\sim 18-19$ mag. As shown in
Figure \ref{MCbrightness}, with this limit the fraction of the
optical afterglows that can be detectable is only $\sim $ 19\%. We
check the detection probability with
ROTSE\footnote{http://www.rotse.net/grb\_reports/} and find that
this faction is 27/89 for the ROTSE responses earlier than $10^3$
seconds post the GRB triggers. The slight difference between our
prediction and the ROTSE observations may be due to the fact that
the ROTSE responses are usually much earlier than $10^3$ s post the
GRB triggers. The current largest robotic optical telescopes for GRB
optical afterglow survey are the Liverpool and Faulkes Telescopes
(North and South). They have a limit of $\sim 22$ mag. The detection
fraction with these telescopes is 24/63 in the first 10 minutes.
This fractions are consistent with that derived from  Figure
\ref{MCbrightness} with a limit of $M_{\rm R}=22$ mag, i.e., 37\%.
The UV-optical telescope (UVOT) on board {\em Swift}, which has a
limit of $\sim 20$ mag, promptly slewed to the GRB trigger positions
and detects $\sim 27\%$ of GRBs at the $3\sigma$ level (in an
individual exposure)\footnote{Note that the fraction of UV-optical
counterpart detection with UVOT for {\rm Swift} GRBs varies for the
epoch selection and the limit from one exposure or co-add more
exposure (Roming et al. 2006; 2009).}. The fraction of UVOT
detection is consistent with our result shown in Figure
\ref{MCbrightness}. Note that, the detection probability with the
Palomar 60 inch telescope, which is a robotic telescope with a
response time of $\sim 180$ seconds and a limiting magnitude of
$M_{\rm R}=20.5$ (Cenko et al. 2006), is 22/29 from a sample made
during 2005-2008 (Cenko et al. 2009). The high detection efficiency
may be due to the fact that P60 has a redder coverage than UVOT. The
detection efficiency of GROND, a simultaneous 7-channel
optical/near-infrared imager (Greiner et al. 2008) mounted at the
2.2m MPI/ESO telescope with a detection limit of $M_{\rm R}\sim
25.3$ mag, is even higher, which is $ \sim 91\%$ and $\sim 88\%$
within 0.5 hours and $0.5\sim 4 $ hours after the trigger,
respectively (Greiner et al. 2011). The fraction of GRBs with
$M_{\rm R}=25.3$ mag derived from Figure \ref{MCbrightness} is
$77\%$, which is slightly lower than the detection probability with
GROND. This would be also due to redder coverage of GROND.

\section{CONCLUSIONS AND DISCUSSION\label{sec:normal}}
We have presented the brightness distributions of early optical
afterglows at different epochs derived from the empirical fits to
the observed lightcurves for a sample of 93 GRBs that have optical
detection at $t<1$ hour post the GRB triggers. The typical R-band flux
at $t=10^2$ s, $10^3$s, and 1 hour are 16.1, 17.1, and 18.4 mag, respectively,
for the all sample. We also derive the distributions at these epochs for the sub-samples that show an
afterglow onset bump feature and the shallow decay segment. They are generally
consistent with that derived from the global sample. The brightness at the
peak time of the bumps falls in the range of $R=9\sim 22$ mag, with a typical
value of $M_{\rm R}=15.6$ mag. The typical brightness at the break
time of the shallow decay segment is $M_{\rm R}=18.3$ mag.

We further perform Monte Carlo simulations to study the luminosity
functions of the optical afterglows for the global sample at the rest-frame
$10^3$ seconds, for the afterglow onset bumps at $t_{\rm p}$, and
for the shallow decay segment at $t_{\rm b}$. We find that
a single power-law luminosity function is adequate to describe the
data, with the indices $-1.40\pm 0.10$, $-1.06\pm 0.10$, and $-1.54\pm
0.22$ for the three samples, respectively. Based on the derived luminosity
function, we generate the intrinsic distribution of $M_{\rm R}$ at $10^{3}$
seconds and show that the distribution peaks at $M_{\rm R}=22.5$ mag.
The detection probabilities of the optical afterglows with
ground-based robotic telescopes and UVOT onboard {\em Swift} are
consistent with our results.

\begin{figure*}
\includegraphics[angle=0,scale=1]{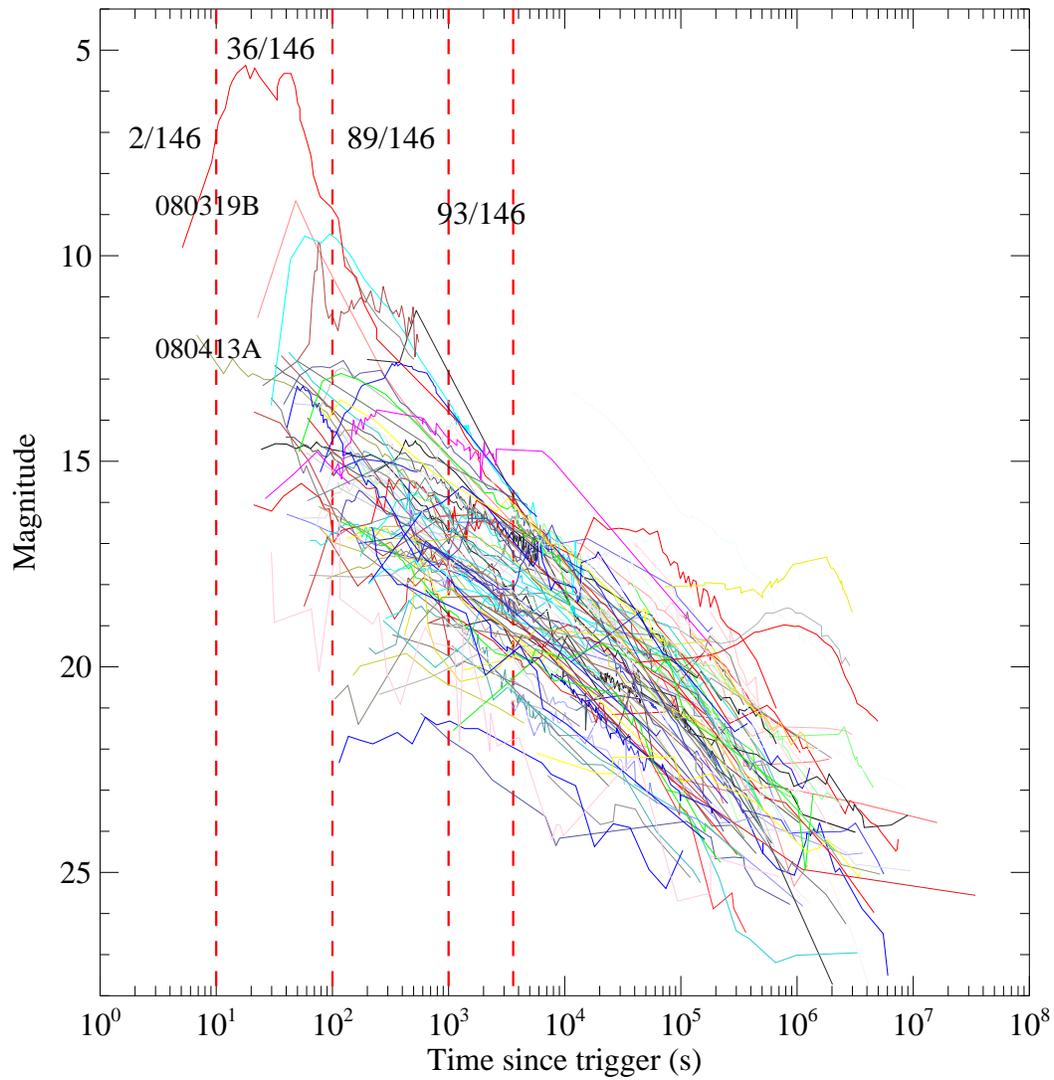}
\caption{Galactic extinction corrected R band lightcurves of 146 GRBs. Different colors are used to make it easier to discern different GRBs. The detection fractions at different epochs are marked.}\label{total}
\end{figure*}

The nature of optically dark GRBs is still not fully uncovered. The proposed
explanations for the dark GRBs are essentially classified into two
classes: extrinsic and intrinsic effects. The extrinsic effects include
extinction and/or absorption by the host galaxies, foreground
extinction, Lyman-$\alpha$ absorption, etc. The intrinsic effects include
intrinsic faintness or rapidly decay of the optical
afterglow due to a low-density environment, and suppression of reverse shock
optical emission in a Poynting-flux dominated jet, etc (Zhang 2007
and references therein). Some authors quantified the degree of
optical darkness with the upper limit on the afterglow flux (Rol et
al. 2005) or with the optical-to-X-ray spectral index (Jakobsson et
al. 2004). Note that the faction of GRBs with no optical detection is
dramatically different among samples, ranging from $20\%$ to $60\%$ (Melandri et al. 2008, 2012;  Cenko et al.
2009; Zheng et al. 2009; Gehrels et al. 2008; Fynbo et al. 2009 and
Greiner et al. 2011). The non-detection of an optical counterpart in these samples should be partially due to observation selection effects. For example, about $\sim 27\%$ of {\em Swift} GRBs were
detected with UVOT at the $3\sigma$ level (in an individual
exposure; Roming et al. 2009), but $\sim 60\%$ were detected by
ground-based telescopes in redder bands than the UVOT bands. Although we cannot exclude the existence of an intrinsically optically dark GRB population, the intrinsic $M_{\rm R}$ distribution derived from our luminosity functions at $t=10^{3}$ seconds may, at least partially, account for the non-detection of an optical counterpart in some GRBs. As discussed in \S 2, the non-detection of optical afterglows dimmer than 22 mag at $t=10^3$ seconds in our sample would be due to the observational selection effect. From
Figure \ref{total}, one can see that the fraction of GRBs with optical afterglows dimmer than 22 at this epoch is
$\sim 63\%$. Even with a deep limit of $M_{\rm R}=25$ mag, this fraction
is still $\sim 25\%$. Therefore, the variation of the fractions of optically dark GRBs may be partially due to the sample selection effect.

\begin{figure*}
\includegraphics[angle=0,scale=0.6]{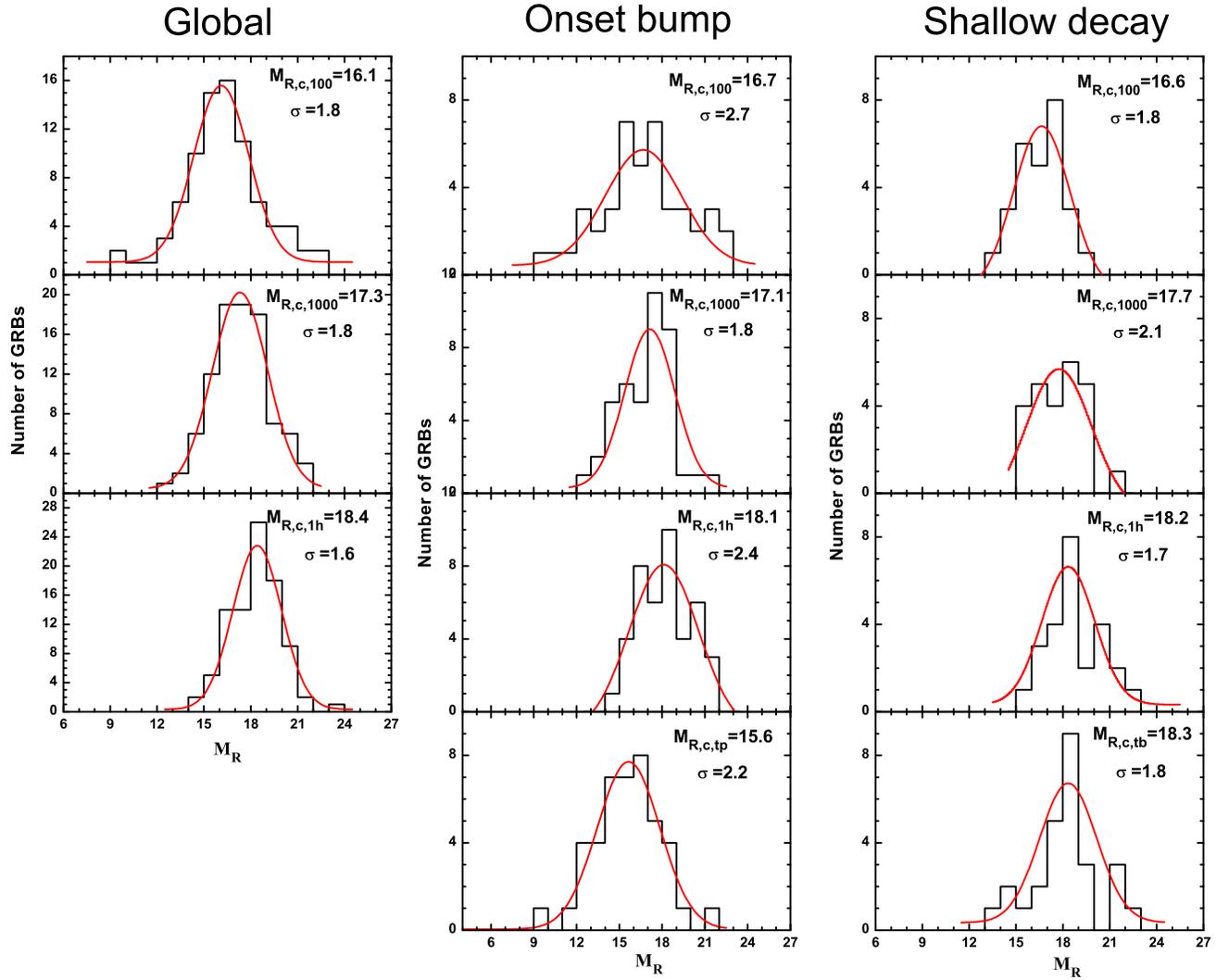}
\caption{R band magnitude distributions with Gaussian fits at
different epochs for the full sample (left panels), the
onset bump sample (middle panels), and the shallow decay segment
sample (right panels). The central values of the Gaussian fits are also
marked.}\label{Dis_obs}
\end{figure*}

\begin{figure*}
\includegraphics[angle=0,scale=0.4]{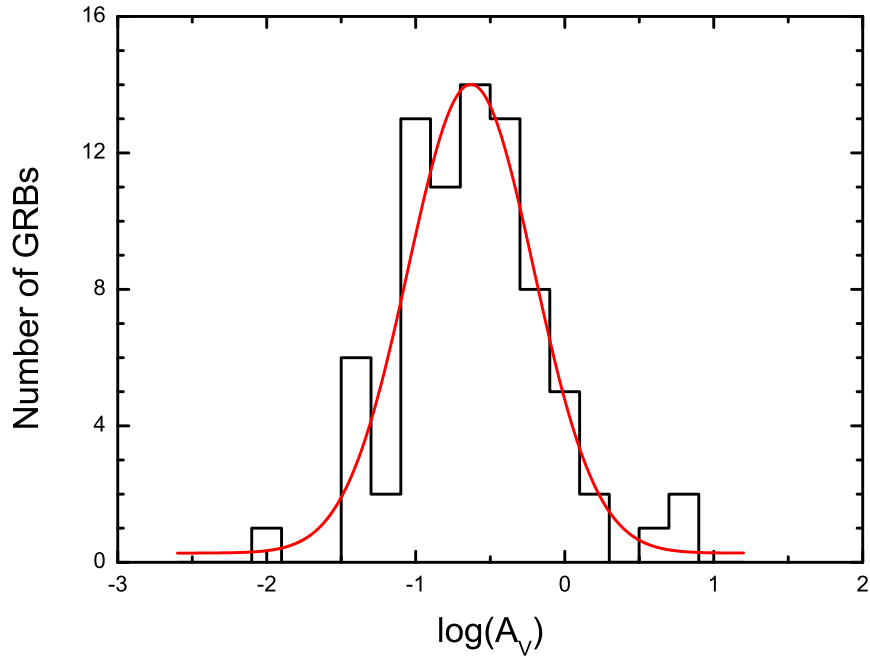}
\caption{Distribution of $\log A_{V}$ values along with our best fit
with a Gaussian function for 79 GRBs collected from the literature. The
Gaussian fit gives $\log(\emph{A}_{V})=-0.63\pm0.42$(1$\sigma$).}\label{Av}
\end{figure*}

\begin{figure*}
\includegraphics[angle=0,scale=0.6]{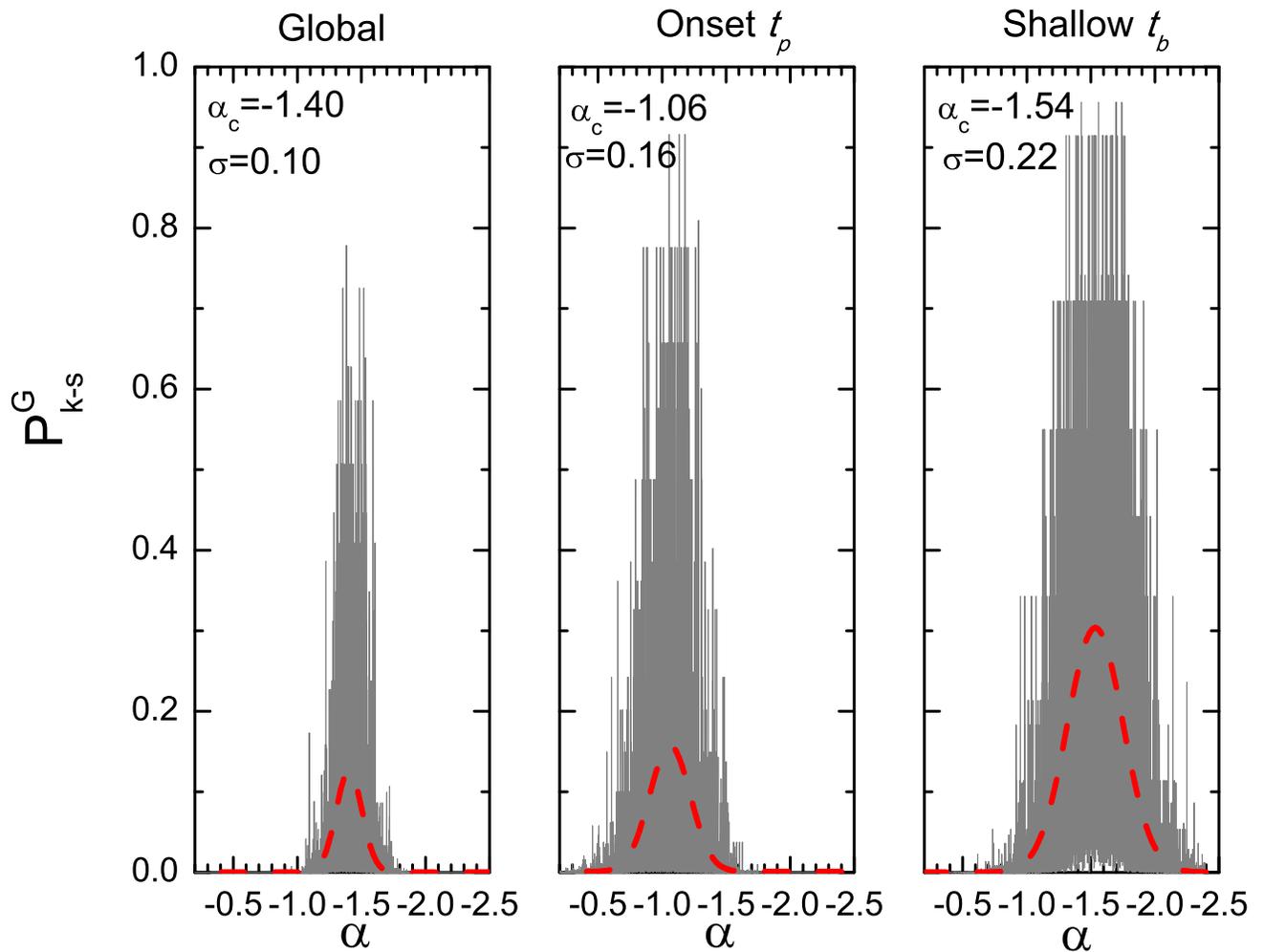}
\caption{Distributions of the K-S test probability $P_{\rm K-S}^{G}$
of the power-law indices of the optical luminosity functions for the
full sample at $t=10^3$ seconds (panel a), for the onset bump sample at
$t_{\rm p}$ (panel b), and for the shallow decay segment sample at $t_{\rm
b}$ (panel c). The dashed curves are Gaussian fits to the distributions
of the average $P_{\rm K-S}^{G}$ in each bin. The central values of
the Gaussian fits are also marked.}\label{Pks_alpha_dis}
\end{figure*}

\begin{figure*}
\includegraphics[angle=0,scale=0.5]{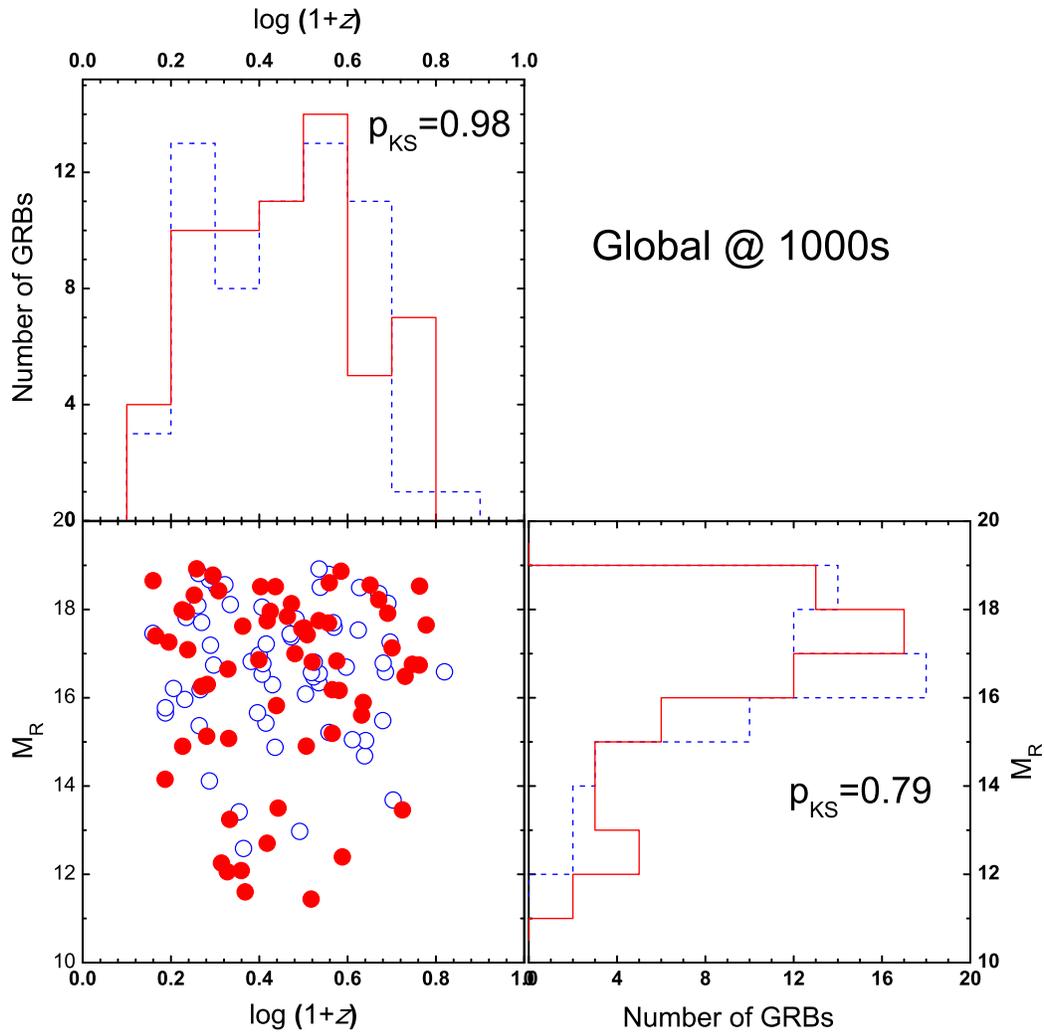}
\caption{Illustration of the best consistency between the observations (open dots and
dashed lines) and our simulations with an optical luminosity function $\Phi\propto L^{-1.40\pm 0.10}$ (solid dots and solid lines) for the full sample at $t=10^{3}$s in two-dimensional $\log M_{\rm R}-\log (1+z)$
distributions and one-dimensional $\log (1+z)$ and $M_{\rm R}$
distributions.}\label{MCtotal1000}
\end{figure*}

\begin{figure*}
\includegraphics[angle=0,scale=0.5]{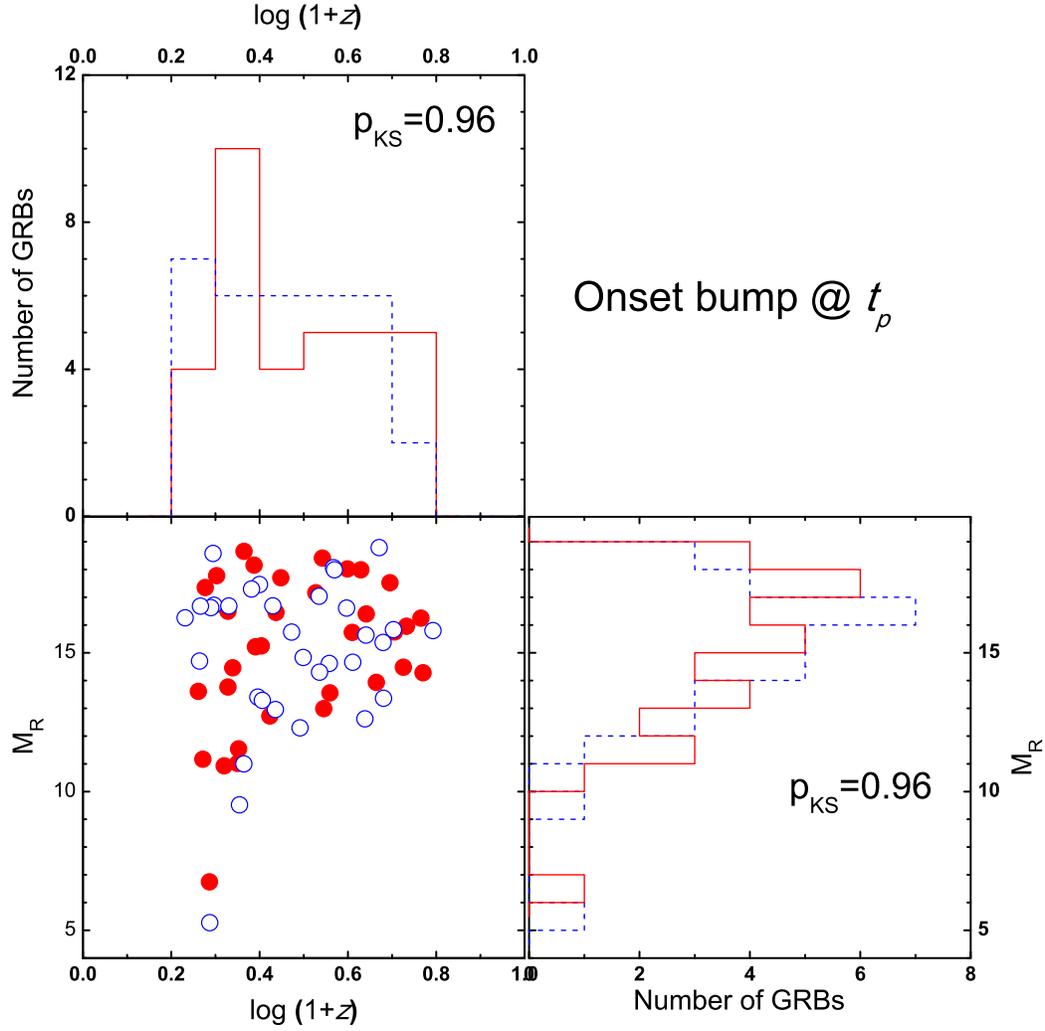}
\caption{Same as Fig. \ref{MCtotal1000}, but for the onset bump
sample at $t_p$ with a luminosity function $\Phi\propto
L^{-1.06\pm 0.16}$.}\label{MConset}
\end{figure*}

\begin{figure*}
\includegraphics[angle=0,scale=0.5]{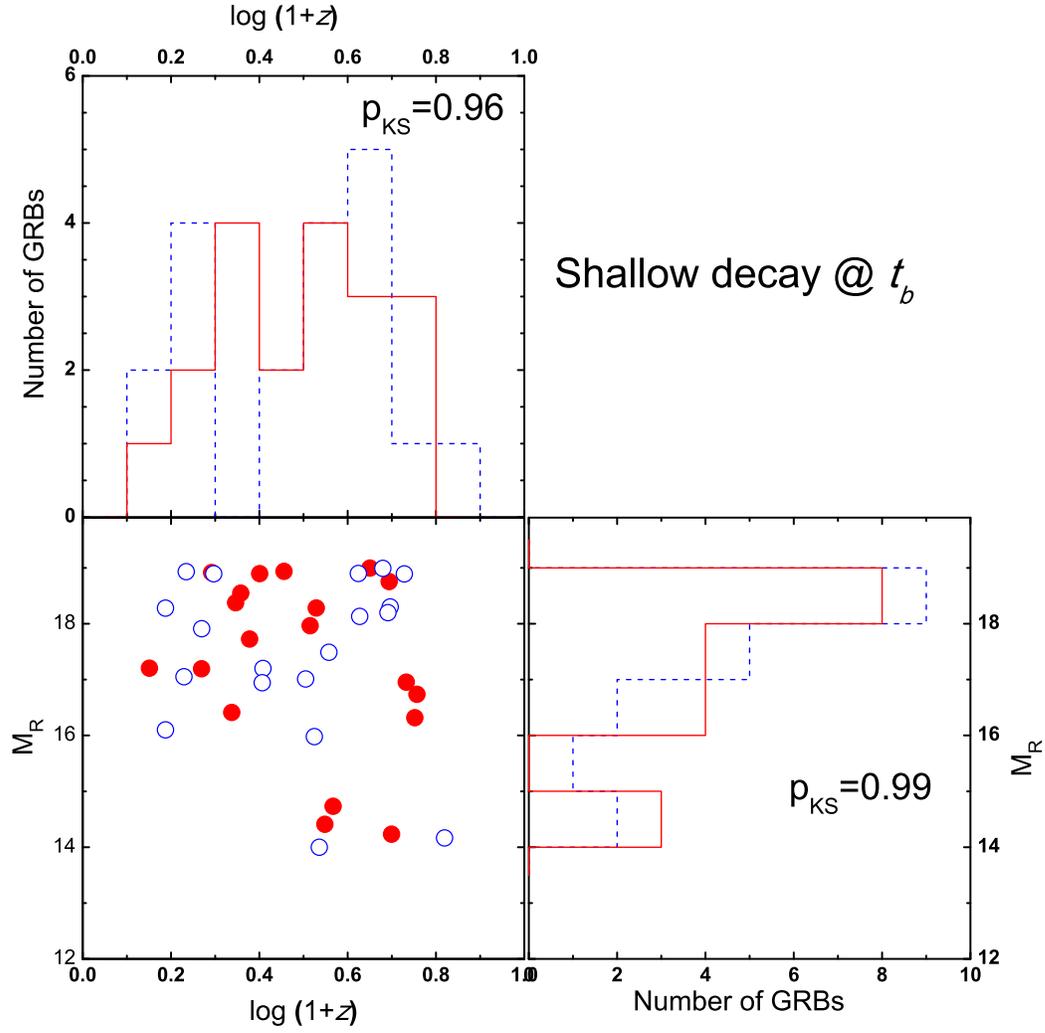}
\caption{Same as Fig. \ref{MCtotal1000}, but for the shallow decay segment at $t_{\rm b}$ with a
luminosity function $\Phi\propto L^{-1.54\pm
0.22}$.}\label{MCshallow}
\end{figure*}

\begin{figure*}
\includegraphics[angle=0,scale=0.5]{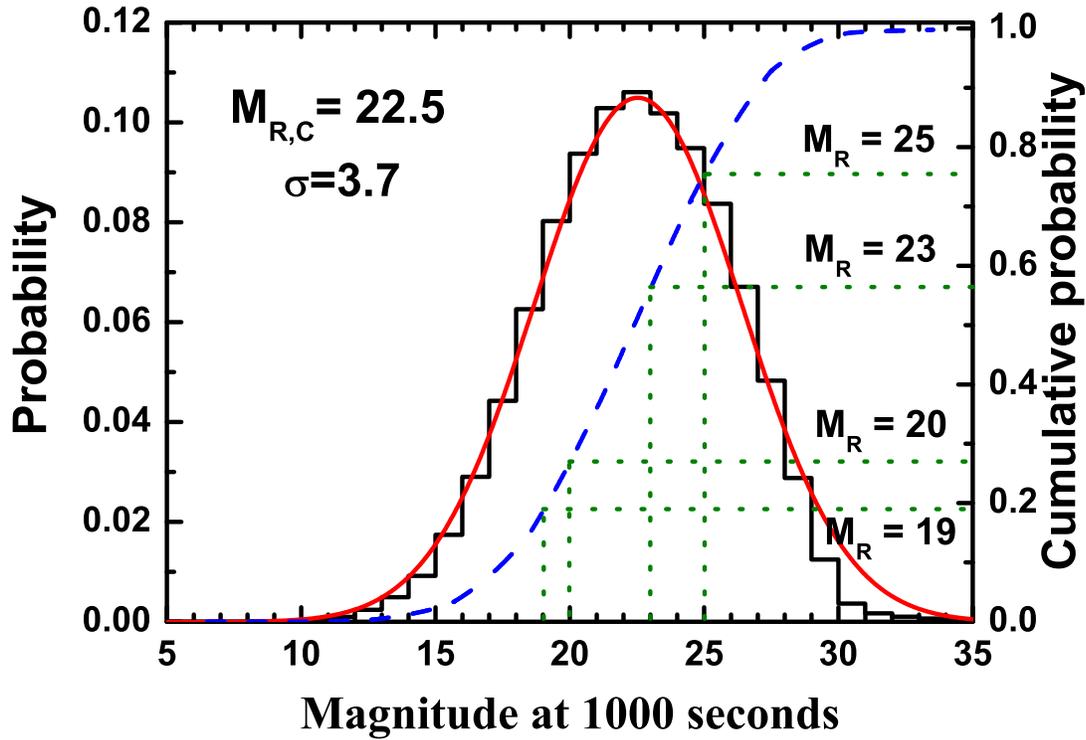}
\caption{The observer-frame probability distributions of the apparent R band magnitude at
$t=10^{3}$ seconds (histogram) along with the Gaussian fit (smooth
curve). The cumulative probability distribution is also shown with a
dashed line. The peak of the probability distribution is $M_{\rm R,
c}=22.5$ mag. The dotted lines mark the probabilities of the optical
afterglow detection at the instrument thresholds $M_{\rm R}=19$ and
$M_{\rm R}=23$, which correspond to small robotic telescopes and the
VT onboard the future mission SVOM.}\label{MCbrightness}
\end{figure*}

The Visible Telescope (VT) on board the upcoming Space-based Variable Objects Monitor (SVOM), a new Chinese-French mission aiming at studying GRBs, has a limiting magnitude of 23 mag ($5\sigma$ significance level) for a 300 s
exposure time (Paul et al. 2011). The fraction of GRBs that would be detectable with VT inferred from our result is 56\% at $t=10^3$ seconds. As shown in Figure \ref{Dis_obs}, the optical afterglows at $t=10^2$ seconds are brighter than that at $10^{3}$ seconds by about one magnitude. The rapid slewing capacity of VT may further increase this fraction at an earlier epoch. Such a sensitivity is a
significant improvement over the UVOT for the GRB early optical afterglow study.

\acknowledgments
This work is supported by the National Basic Research Program (973 Programme) of China (Grant
2009CB824800), the National Natural Science Foundation of China
(Grants 11025313, 11163001, 11203008), Special Foundation for Distinguished Expert Program of Guangxi, and the Guangxi Natural Science Foundation (2013GXNSFFA019001, 2010GXNSFA013112, 2010GXNSFC013011, and Contract No. 2011-135). BZ acknowledges support from NSF (AST-0908362).


\clearpage
\begin{deluxetable}{llllllllll}
\tablewidth{500pt} 
\tablecaption{The temporal coverage of the lightcurves, redshifts, spectral indices, apparent magnitudes at selected epochs for 93 GRBs that have well-sampled early optical lightcurves}
\tablenum{1} \tablehead{ \colhead{GRB}  &\colhead{$T_{\rm start}$(s)}
&\colhead{$T_{\rm end}$(s)} &\colhead{$z$} &\colhead{$\beta_{\rm O}$}
 &\colhead{$M_{\rm R, 10^2{\rm s}}$} &\colhead{$M_{\rm R, 10^3{\rm s}}$} &\colhead{$M_{\rm R, 1{\rm h}}$}}
\startdata

990123  &   22  &   1229240    &   1.6 &   0.75    $\pm$   0.07    &   10.2    &   15.4    &   17.3    \\
021004  &   155 &   2421620 &   2.335   &   0.39            &   15.4    &   16.1    &   16.5    \\
021211  &   130 &   8957    &   1.01    &   0.69            &   14.1    &   17.8    &   19.3    \\
030418  &   288 &   7192    &   ... &   ...         &   18.3    &   16.5    &   16.8    \\
040924  &   950 &   62986   &   0.859   &   0.70            &   17.5    &   17.7    &   18.7    \\
041006  &   230 &   5577258 &   0.716   &   0.55            &   16.5    &   17.6    &   18.2    \\
041219A &   441 &   186558  &   ... &   ...         &   ... &   17.2    &   16.9    \\
050319  &   40  &   994118  &   3.24    &   0.74    $\pm$   0.42    &   16.7    &   18.1    &   19.0    \\
050401  &   3456    &   1120000 &   2.9 &   0.39    $\pm$   0.05    &   ... &   19.9    &   20.9    \\
050408  &   3352    &   36710070    &   1.2357  &   0.28    $\pm$   0.33    &   18.3    &   19.6    &   20.3    \\
050502A &   47  &   17850   &   3.793   &   0.76    $\pm$   0.16    &   13.9    &   17.0    &   18.3    \\
050525A &   66  &   35638   &   0.606   &   0.97    $\pm$   0.10    &   14.2    &   16.1    &   17.8    \\
050721  &   1484    &   248596  &   ... &   1.16    $\pm$   0.35    &   ... &   16.8    &   18.6    \\
050730  &   67  &   72697   &   3.969   &   0.52    $\pm$   0.05    &   15.9    &   16.7    &   17.3    \\
050801  &   24  &   21652   &   1.56    &   1.00    $\pm$   0.16    &   14.7    &   16.7    &   18.1    \\
050820A &   230 &   663300  &   2.612   &   0.72    $\pm$   0.03    &   16.5    &   15.2    &   16.7    \\
050824  &   635 &   8994457 &   0.83    &   0.40    $\pm$   0.04    &   17.1    &   18.5    &   19.2    \\
050922C &   745 &   606010  &   2.198   &   0.51    $\pm$   0.05    &   14.9    &   15.9    &   16.9    \\
051021  &   1611    &   35820   &   ... &   0.75            &   17.5    &   18.2    &   18.6    \\
051109A &   40  &   1040000 &   2.346   &   0.70            &   14.9    &   16.6    &   17.7    \\
051111  &   32  &   7588    &   1.55    &   0.76    $\pm$   0.07    &   13.7    &   15.7    &   17.2    \\
060110  &   25  &   4781    &   ... &   0.75            &   12.9    &   14.8    &   15.9    \\
060111B &   30  &   264806  &   ... &   0.70            &   16.2    &   19.0    &   20.6    \\
060124  &   3335    &   1980000 &   2.296   &   0.73    $\pm$   0.08    &   ... &   15.4    &   16.7    \\
060206  &   319 &   201580  &   4.048   &   0.73    $\pm$   0.05    &   15.5    &   17.9    &   19.7    \\
060210  &   63  &   7190    &   3.91    &   0.37            &   17.8    &   18.1    &   19.7    \\
060218  &   253 &   2850000 &   0.0331  &   0.75            &   21.0    &   20.1    &   19.6    \\
060418  &   76  &   7659    &   1.489   &   0.78    $\pm$   0.09    &   13.9    &   15.7    &   17.4    \\
060512  &   112 &   5927    &   0.4428  &   0.68    $\pm$   0.05    &   15.4    &   17.2    &   18.2    \\
060526  &   60  &   893550  &   3.21    &   0.51    $\pm$   0.32    &   15.0    &   16.5    &   17.3    \\
060605  &   74  &   6317    &   3.78    &   1.06            &   15.9    &   15.6    &   17.1    \\
060607A &   73  &   14733   &   3.082   &   0.72    $\pm$   0.27    &   15.9    &   17.0    &   19.0    \\
060614  &   1547    &   1276350 &   0.125   &   0.47    $\pm$   0.04    &   20.9    &   20.1    &   19.6    \\
060729  &   696 &   662390  &   0.54    &   0.78    $\pm$   0.03    &   15.3    &   15.6    &   15.8    \\
060904B &   21  &   163131  &   0.703   &   1.11    $\pm$   0.10    &   16.8    &   16.4    &   18.9    \\
060906  &   661 &   13610   &   3.686   &   0.56    $\pm$   0.02    &   20.1    &   18.8    &   19.8    \\
060908  &   825 &   7242    &   2.43    &   0.30            &   14.7    &   17.5    &   19.0    \\
060912A &   1100    &   23900   &   0.937   &   0.62            &   ... &   18.0    &   19.3    \\
060926  &   57  &   1200    &   3.2 &   0.82    $\pm$   0.01    &   16.7    &   19.1    &   ... \\
060927  &   17  &   1169    &   5.6 &   0.86    $\pm$   0.03    &   16.0    &   16.6    &   18.2    \\
061007  &   30  &   14599   &   1.261   &   0.78    $\pm$   0.02    &   9.6 &   13.4    &   15.7    \\
061121  &   49  &   554 &   1.1588  &   0.95            &   11.6    &   12.6    &   14.1    \\
061126  &   36  &   156381  &   1.1588  &   0.95            &   13.8    &   17.3    &   18.5    \\
070110  &   662 &   34762   &   1.547   &   0.55    $\pm$   0.04    &   18.6    &   19.0    &   19.2    \\
070208  &   329 &   13460   &   1.165   &   0.68            &   18.6    &   19.8    &   20.4    \\
070311  &   74  &   350926  &   ... &   1.00    $\pm$   0.20    &   14.8    &   16.6    &   17.6    \\
070318  &   61  &   87366   &   0.836   &   0.78            &   15.5    &   15.4    &   17.0    \\
070411  &   184 &   516628  &   2.954   &   0.75            &   16.5    &   16.6    &   18.1    \\
070419A &   206 &   62218   &   0.97    &   0.80            &   20.4    &   18.8    &   20.5    \\
070420  &   116 &   10842   &   ... &   0.75            &   15.6    &   16.2    &   17.4    \\
070518  &   1069    &   311763  &   1.16    &   0.80            &   18.0    &   19.6    &   20.6    \\
070611  &   274 &   8867    &   2.04    &   0.73            &   17.7    &   20.1    &   18.6    \\
071003  &   569 &   5003    &   1.605   &   1.25    $\pm$   0.09    &   13.4    &   17.3    &   18.7    \\
071010A &   321 &   523226  &   0.98    &   0.68            &   17.0    &   16.7    &   17.7    \\
071010B &   64  &   174464  &   0.947   &   0.00            &   16.8    &   17.2    &   18.0    \\
071025  &   175 &   14885   &   ... &   0.42    $\pm$   0.08    &   17.7    &   16.2    &   17.5    \\
071031  &   287 &   350926  &   2.692   &   0.64    $\pm$   0.01    &   19.5    &   18.1    &   18.8    \\
071112C &   132 &   69638   &   0.823   &   0.63    $\pm$   0.29    &   17.5    &   18.1    &   19.3    \\
071122  &   1303    &   9047    &   1.14    &   0.83            &   ... &   20.1    &   19.8    \\
080310  &   151 &   124416  &   2.4266  &   0.42    $\pm$   0.12    &   17.3    &   16.9    &   17.5    \\
080319A &   150 &   4462    &   ... &   0.77    $\pm$   0.02    &   21.0    &   20.3    &   21.2    \\
080319B &   5   &   4590000 &   0.937   &   0.75            &   9.4 &   14.1    &   16.0    \\
080319C &   78  &   1432    &   1.949   &   0.77    $\pm$   0.02    &   16.8    &   17.4    &   18.7    \\
080330  &   89  &   116557  &   1.51    &   0.49            &   17.8    &   17.5    &   18.3    \\
080413A &   7   &   18339   &   2.433   &   0.67            &   14.1    &   16.5    &   18.2    \\
080413B &   77  &   5185072 &   1.1 &   0.25    $\pm$   0.07    &   16.2    &   18.1    &   18.9    \\
080506  &   210 &   5371    &   ... &   0.95    $\pm$   0.05    &   16.4    &   17.6    &   18.5    \\
080603A &   105 &   350436  &   ... &   0.75            &   21.5    &   18.4    &   18.5    \\
080710  &   417 &   34762   &   0.845   &   0.80    $\pm$   0.09    &   19.1    &   16.9    &   16.7    \\
080804  &   1160    &   26112   &   2.2 &   0.43            &   ... &   17.9    &   19.2    \\
080810  &   38  &   7898    &   3.35    &   0.44            &   12.6    &   14.7    &   16.4    \\
080913  &   576 &   870036  &   6.7 &   0.79    $\pm$   0.03    &   19.3    &   21.7    &   23.1    \\
080928  &   390 &   13425   &   1.692   &   1.08    $\pm$   0.02    &   18.7    &   17.0    &   17.2    \\
081008  &   109 &   184525  &   1.692   &   1.08    $\pm$   0.02    &   17.0    &   17.4    &   18.7    \\
081029  &   529 &   252674  &   3.85    &   0.00            &   15.6    &   17.4    &   18.0    \\
081109A &   169 &   66600   &   ... &   0.75            &   17.8    &   18.1    &   19.3    \\
081126  &   102 &   541 &   ... &   0.75            &   15.4    &   15.5    &   16.1    \\
081203A &   78  &   5758    &   2.1 &   0.60            &   14.0    &   13.1    &   15.2    \\
090102  &   41  &   264553  &   1.547   &   0.74            &   14.2    &   18.1    &   19.7    \\
090313  &   205 &   7874700 &   3.375   &   0.71            &   18.8    &   15.8    &   16.8    \\
090426  &   86  &   10748   &   2.609   &   0.76    $\pm$   0.14    &   16.4    &   18.6    &   19.8    \\
090510  &   114 &   103794  &   0.903   &   0.76    $\pm$   0.14    &   22.2    &   21.5    &   21.9    \\
090618  &   76  &   72576   &   0.54    &   0.50            &   13.7    &   15.5    &   16.4    \\
090726  &   204 &   3015    &   2.71    &   0.75            &   19.1    &   20.7    &   ... \\
090812  &   27  &   142 &   2.452   &   0.36            &   15.1    &   18.5    &   20.4    \\
100219A &   936 &   34978   &   0.49    &   0.18            &   16.4    &   19.0    &   20.8    \\
100418A &   1099    &   1371570 &   ... &   0.75            &   ... &   21.4    &   20.5    \\
100728B &   159 &   5644    &   ... &   0.75            &   16.3    &   18.8    &   19.7    \\
100901A &   634 &   543008  &   1.408   &   0.75            &   22.2    &   17.6    &   18.2    \\
100906A &   51  &   10937   &   1.408   &   0.75            &   13.0    &   14.9    &   16.4    \\
101024A &   219 &   160000  &   ... &   0.75            &   ... &   19.3    &   19.8    \\
110205A &   540 &   384000  &   ... &   0.75            &   ... &   14.2    &   16.1    \\
110213A &   104 &   183368  &   ... &   0.75            &   15.1 & 14.4    &   14.8
\enddata
\end{deluxetable}

\begin{deluxetable}{llllllllll}
\tablewidth{400pt} 
\tabletypesize{\footnotesize}
\tablecaption{The temporal coverage of the lightcurves, redshifts, spectral indices, peak time, apparent magnitudes at selected epochs for 39 GRBs that have an onset bump in their optical lightcurves}
\tablenum{2} \tablehead{\colhead{GRB} &\colhead{$z$}
&\colhead{$T_{\rm start}$(s)} &\colhead{$T_{\rm end}$(s)}
&\colhead{$t_{\rm p}$(s)} &\colhead{$M_{\rm R, 10^2{\rm s}}$}
&\colhead{$M_{\rm R, 10^3{\rm s}}$} &\colhead{$M_{\rm R, 1{\rm h}}$}
&\colhead{$M_{\rm R, t_{\rm p}}$}}
\startdata

030418  &   ... &   288 &   7192    &   1190    &   18.3    &   16.5    &   16.8    &   16.4    \\
050502A &   3.793   &   47  &   17850   &   58  &   13.9    &   17.1    &   19.0    &   13.4    \\
050820A &   2.612   &   230 &   663300  &   477 &   16.5    &   15.2    &   16.7    &   14.6    \\
060110  &   ... &   25  &   4781    &   50  &   12.9    &   14.8    &   15.9    &   12.5    \\
060111B &   ... &   30  &   264806  &   54  &   16.2    &   19.0    &   20.6    &   15.7    \\
060418  &   1.489   &   76  &   7659    &   170 &   13.9    &   15.7    &   17.4    &   13.4    \\
060605  &   3.78    &   74  &   6317    &   590 &   16.1    &   15.7    &   17.5    &   15.4    \\
060607A &   3.082   &   73  &   14733   &   179 &   15.9    &   17.0    &   19.0    &   14.7    \\
060904B &   0.703   &   21  &   163131  &   85  &   17.2    &   ... &   ... &   16.3    \\
060906  &   3.686   &   661 &   13610   &   1149    &   20.1    &   18.8    &   20.0    &   18.8    \\
061007  &   1.261   &   30  &   14599   &   77  &   9.6     &   13.4    &   15.7    &   9.5     \\
061121  &   1.1588  &   49  &   554     &   208 &   11.7    &   12.6    &   14.1    &   11.0    \\
070318  &   0.836   &   61  &   87366   &   507 &   15.5    &   15.4    &   17.0    &   14.7    \\
070419A &   0.97    &   206 &   62218   &   765 &   20.4    &   18.8    &   20.5    &   18.6    \\
070420  &   ... &   116 &   10842   &   202 &   15.6    &   16.2    &   17.4    &   14.8    \\
071010A &   0.98    &   321 &   523226  &   586 &   19.3    &   17.1    &   18.7    &   16.7    \\
071010B &   0.947   &   64  &   174464  &   287 &   16.8    &   17.2    &   18.0    &   16.6    \\
071025  &   ... &   175 &   14885   &   548 &   17.7    &   16.2    &   17.7    &   15.8    \\
071031  &   2.692   &   287 &   350926  &   1213    &   19.5    &   18.1    &   18.8    &   18.1    \\
071112C &   0.823   &   132 &   69638   &   178 &   17.5    &   18.1    &   19.3    &   16.7    \\
080310  &   2.4266  &   298 &   124416  &   184 &   17.8    &   19.0    &   20.6    &   17.0    \\
080319A &   ... &   151 &   4462    &   238 &   21.0    &   20.3    &   21.2    &   19.6    \\
080330  &   1.51    &   89  &   116557  &   578 &   17.8    &   17.9    &   19.4    &   17.5    \\
080413A &   2.433   &   7   &   18339   &   150 &   14.5    &   16.5    &   18.2    &   14.3    \\
080603A &   ... &   105 &   350436  &   1044    &   21.5    &   18.4    &   18.5    &   18.3    \\
080710  &   0.845   &   417 &   34762   &   1934    &   21.1    &   17.2    &   17.0    &   16.7    \\
080810  &   3.35    &   38  &   7898    &   117 &   12.6    &   14.7    &   16.4    &   12.6    \\
080928  &   1.692   &   390 &   13425   &   2290    &   18.7    &   17.0    &   17.4    &   16.7    \\
081008  &   1.692   &   109 &   184525  &   163 &   17.0    &   17.4    &   18.7    &   15.7    \\
081109A &   ... &   169 &   66600   &   559 &   17.8    &   18.1    &   19.3    &   17.7    \\
081126  &   ... &   102 &   541     &   159 &   15.4    &   15.5    &   16.1    &   15.0    \\
081203A &   2.1 &   78  &   5758    &   295 &   14.0    &   13.1    &   15.2    &   12.3    \\
090102  &   1.547   &   41  &   264553  &   50  &   14.2    &   18.1    &   20.3    &   13.3    \\
090313  &   3.375   &   205 &   7874700     &   1315    &   18.8    &   15.8    &   16.8    &   15.6    \\
090510  &   0.903   &   114 &   103794  &   1579    &   22.2    &   21.5    &   21.9    &   21.5    \\
090726  &   2.71    &   204 &   3015    &   290 &   19.2    &   ... &   ... &   18.0    \\
090812  &   2.452   &   27  &   142     &   71  &   15.1    &   18.5    &   20.4    &   14.8    \\
100901A &   1.408   &   634 &   543008  &   1260    &   22.2    &   17.6    &   18.2    &   17.3    \\
100906A &   1.408   &   51  &   10937   &   101 &   13.0    &   14.9    &   16.4    &   13.0    \\
110205A &   ... &   540 &   384000  &   948 &   ... &   14.2    &   16.1    &   14.2    \\
110213A &   ... &   104 &   183368  &   294 &   15.1    &   14.6 &
15.8    &   13.6
\enddata
\end{deluxetable}
\begin{deluxetable}{llllllllll}

\tablewidth{400pt} 
\tabletypesize{\footnotesize}
\tablecaption{The temporal coverage of the lightcurves, redshifts, spectral indices, peak time, apparent magnitudes at selected epoches for 26 GRBs that have an early shallow decay segment in their optical lightcurves} \tablenum{3} \tablehead{ \colhead{GRB} &\colhead{$z$}
&\colhead{$T_{\rm start}$(s)} &\colhead{$T_{\rm end}$(s)}
&\colhead{$T_{\rm p}$(s)} &\colhead{$M_{\rm R, 10^2{\rm s}}$}
&\colhead{$M_{\rm R, 10^3{\rm s}}$} &\colhead{$M_{\rm R, 1{\rm h}}$}
&\colhead{$M_{\rm R, t_{\rm b}}$}}

\startdata
021004  &   2.335   &   156     &   2421620     &   9909    &   15.4    &   16.1    &   16.5    &   17.1    \\
040924  &   0.859   &   950     &   62986   &   1722    &   17.5    &   17.7    &   18.7    &   17.9    \\
041006  &   0.716   &   230     &   5580000     &   11346   &   16.5    &   17.6    &   18.2    &   18.9    \\
050319  &   3.24    &   40  &   994118  &   535     &   17.3    &   19.3    &   22.2    &   18.1    \\
050408  &   1.2357  &   3352    &   3670000     &   40389   &   18.3    &   19.6    &   20.3    &   22.0    \\
050730  &   3.969   &   67  &   72697   &   11362   &   15.9    &   16.7    &   17.3    &   18.3    \\
050801  &   1.56    &   24  &   21652   &   1512    &   16.2    &   16.9    &   18.1    &   17.2    \\
050922C &   2.198   &   745     &   606010  &   4083    &   14.9    &   15.9    &   16.9    &   17.0    \\
051021  &    ...    &   1611    &   35820   &   4932    &   17.5    &   18.2    &   18.6    &   18.8    \\
051109A &   2.346   &   40  &   1040000     &   445     &   14.9    &   16.6    &   17.8    &   16.0    \\
060111B &    ...    &   30  &   13700000    &   36  &   19.5    &    ...    &    ...    &   13.9    \\
060210  &   3.91    &   63  &   7190    &   958     &   17.8    &   18.3    &   19.7    &   18.2    \\
060526  &   3.21    &   60  &   893550  &   20940   &   15.0    &   16.5    &   17.3    &   18.9    \\
060605  &   3.78    &   74  &   6317    &   23275   &   17.7    &   18.2    &   18.4    &   19.0    \\
060729  &   0.54    &   696     &   662390  &   4131    &   15.3    &   15.6    &   15.9    &   16.1    \\
060927  &   5.6 &   17  &   1169    &   37  &   16.1    &   21.6    &   24.6    &   14.2    \\
061126  &   1.1588  &   36  &   156381  &   14809   &   17.1    &   18.1    &   18.7    &   19.6    \\
070110  &   1.547   &   662     &   34762   &   14290   &   18.6    &   19.0    &   19.2    &   19.7    \\
070208  &   1.165   &   1166    &   4854    &   12732   &   18.6    &   19.8    &   20.4    &   21.3    \\
070411  &   2.954   &   183     &   516628  &   108994  &   16.9    &   18.2    &   18.9    &   21.4    \\
070518  &   1.16    &   2113    &   311763  &   30000   &   18.0    &   19.6    &   20.6    &   22.7    \\
071010A &   0.98    &   321     &   523226  &   12528   &   17.1    &   17.8    &   18.2    &   18.9    \\
080413A &   2.433   &   7   &   18339   &   73  &   15.4    &    ...    &    ...    &   14.0    \\
081029  &   3.85    &   529     &   252674  &   2323    &   15.6    &   17.4    &   20.0    &   18.9    \\
090426  &   2.609   &   86  &   10748   &   257     &   16.5    &   19.2    &   21.6    &   17.5    \\
090618  &   0.54    &   76  &   72576   &   31120   &   14.0    & 15.6    &   16.4    &   18.3
\enddata

\end{deluxetable}

\begin{deluxetable}{lllllllll}

\tablewidth{450pt} 
\tabletypesize{\footnotesize}
\tablecaption{Extinction of the GRB host galaxies ($A^{\rm host}_{\rm V}$) collected from the literature}
\tablenum{4}
\tablehead{ \colhead{GRB}& \colhead{$A_{V}$} &\colhead{Ref.}
&\colhead{GRB}& \colhead{$A_{V}$} &\colhead{Ref.} &\colhead{GRB}&
\colhead{$A_{V}$} &\colhead{Ref.}}
\startdata
970228  &   0.50            &   (1) &   050820A &   0.07    $\pm$   0.01    &   (7) &   070419A &   0.42    $\pm$   0.37    &   (3) \\
971214  &   0.43    $\pm$   0.08    &   (1) &   050824  &   0.15    $\pm$   0.03    &   (8) &   070518  &   0.30            &   (3) \\
980425  &   1.90    $\pm$   0.10    &   (2) &   050904  &   1.00            &   (10)    &   071003  &   0.34    $\pm$   0.11    &   (3) \\
980613  &   0.45            &   (-) &   051028  &   0.70            &   (11)    &   071010A &   0.64    $\pm$   0.09    &   (3) \\
980703  &   1.50    $\pm$   0.11    &   (1) &   051111  &   0.20    $\pm$   0.10    &   (-) &   071025  &   1.09    $\pm$   0.20    &   (18)    \\
990510  &   0.22    $\pm$   0.07    &   (3) &   060111B &   3.60    $\pm$   0.50    &   (12)    &   071031  &   0.14    $\pm$   0.13    &   (3) \\
990712  &   0.50    $\pm$   0.10    &   (2) &   060124  &   0.05    $\pm$   0.26    &   (3) &   071112C &   0.23    $\pm$   0.21    &   (3) \\
991208  &   0.05            &   (4) &   060206  &   0.01    $\pm$   0.02    &   (12)    &   080310  &   0.19    $\pm$   0.05    &   (3) \\
000301C &   0.09    $\pm$   0.04    &   (1) &   060210  &   1.18    $\pm$   0.10    &   (3) &   080319C &   0.59    $\pm$   0.12    &   (3) \\
000418  &   0.96            &   (1) &   060418  &   0.12    $\pm$   0.05    &   (2) &   080330  &   0.19    $\pm$   0.08    &   (3) \\
000926  &   0.18    $\pm$   0.06    &   (1) &   060526  &   0.05    $\pm$   0.11    &   (-) &   080413A &   0.13    $\pm$   0.07    &   (3) \\
020813  &   0.14    $\pm$   0.04    &   (1) &   060614  &   0.11    $\pm$   0.03    &   (8) &   080710  &   0.11    $\pm$   0.04    &   (3) \\
021004  &   0.30            &   (1) &   060729  &   0.07    $\pm$   0.02    &   (7) &   080721  &   0.60            &   (19)    \\
030226  &   0.53            &   (1) &   060904B &   0.08    $\pm$   0.08    &   (8) &   080810  &   0.16    $\pm$   0.02    &   (3) \\
030323  &   0.13    $\pm$   0.09    &   (3) &   060906  &   £¼0.09          &   (3) &   080913  &   (0.58)  $\pm$   0.67    &   (3) \\
030328  &   0.05    $\pm$   0.15    &   (5) &   060908  &   0.05    $\pm$   0.03    &   (7) &   080928  &   0.29    $\pm$   0.03    &   (3) \\
030329  &   0.30    $\pm$   0.03    &   (1) &   060912A &   0.46    $\pm$   0.23    &   (14)    &   081008  &   0.29    $\pm$   0.03    &   (8) \\
030429  &   0.34            &   (1) &   060926  &   0.32    $\pm$   0.02    &   (8) &   081203A &   0.09    $\pm$   0.04    &   (3) \\
030723  &   0.32    $\pm$   0.22    &   (5) &   060927  &   £¼0.12          &   (8) &   090102  &   0.12    $\pm$   0.11    &   (3) \\
041219A &   6.80    $\pm$   1.60    &   (6) &   061007  &   0.39    $\pm$   0.01    &   (8) &   090313  &   0.34    $\pm$   0.15    &   (3) \\
050319  &   0.05    $\pm$   0.09    &   (7) &   061126  &   0.10    $\pm$   0.06    &   (3) &   090323  &   0.14    $\pm$   0.00    &   (3) \\
050401  &   0.65    $\pm$   0.04    &   (8) &   070110  &   0.11    $\pm$   0.04    &   (3) &   090328  &   0.22    $\pm$   0.06    &   (20)    \\
050408  &   0.73    $\pm$   0.18    &   (9) &   070125  &   0.11    $\pm$   0.04    &   (15)    &   090618  &   0.30    $\pm$   0.10    &   (21)    \\
050416A &   0.70    $\pm$   0.70    &   (2) &   070306  &   5.45    $\pm$   0.61    &   (16)    &   090902B &   0.20    $\pm$   0.06    &   (22)    \\
050525A &   0.25    $\pm$   0.16    &   (1) &   070311  &   0.80    $\pm$   0.15    &   (17)    &   090926A &   0.13    $\pm$   0.06    &   (3) \\
050730  &   0.12    $\pm$   0.02    &   (8) &   070318  &   0.44    $\pm$   0.11    &   (14)    &   091127  &   0.20            &   (23)    \\
050801  &   0.30    $\pm$   0.18    &   (2) &

\enddata
\tablerefs{ (1) Liang \& Zhang (2006); (2) Mannucci et al.(2011);
(3) Kann et al.(2010); (4) Sokolov et al.(2001);(5) Kann et
al.(2006); (6) Gotz, D. et al (2011) (7) de Ugarte Postigo et
al.(2011); (8) Zafar et al.(2011); (9) de Ugarte Postigo et
al.(2007); (10) Berger et al.(2007); (11) Castro-Tirado (2006); (12)
Fynbo et al.(2009); (13) Ferrero et al.(2009); (14) Schady et
al.(2011);
 (15) Schady et al.(2008);
 (16) Jaunsen et al.(2008);
 (17) Guidorzi  et al.(2007);
 (18) Perley et al.(2009);
 (19) Starling et al.(2009);
 (20) McBreen (2010);
 (21) Cano et al.(2011);
 (22) Pandey et al.(2010);
 (23) Vergani et al.(2011).
  }
\end{deluxetable}

\clearpage
\thispagestyle{empty}
\setlength{\voffset}{-18mm}

\end{document}